\documentclass[preprint,11pt]{aastex}

\def\kms{\relax \ifmmode {\,\rm km\,s}^{-1}\else \,km\,s$^{-1}$\fi}
\def\ergcms{${\rm erg}^{-1}\,{\rm cm}^{-2}\,{\rm s}^{-1}$}
\newcommand{\Mdot}{\dot{M}}  
\def\mlr{\rm M$_\odot$~yr$^{-1}$}
\def\cm3{${\rm cm}^{-3}$}
\def\nii{[N {\sc ii}]}
\def\h2{H$_2$}
\def\H2{\hbox{H$_2$ (1-0) S(1)}}
\def\h212{\hbox{H$_2$ (2-1) S(1)}}
\def\brgamma{Br~$\gamma$}
\def\n2346{\hbox{NGC 2346}}
\newcommand{\beq}{\begin{equation}}
\newcommand{\eeq}{\end{equation}}

\shorttitle{On the Formation of Molecular Clumps in PNe}
\shortauthors{ Manchado et al.}
\begin{document}

\title{High resolution imaging of NGC 2346 with GSAOI/GeMS: disentangling the planetary nebula molecular structure to understand
its origin and evolution.}
\author{Arturo Manchado\altaffilmark{1,2,3}, Letizia Stanghellini\altaffilmark{4}, Eva Villaver\altaffilmark{5}, Guillermo Garc\'{\i}a-Segura\altaffilmark{6}, Richard A.\ Shaw\altaffilmark{4} \& D.\ A.\ Garc\'\i a-Hern\'andez\altaffilmark{1,2}}

\altaffiltext{1}{Instituto de Astrof\'{\i}sica de Canarias, V\'{\i}a L\'actea S/N, E--38205 La Laguna, Tenerife, Spain}
\altaffiltext{2}{Departmento de Astrof\'{\i}sica, Universidad de La Laguna (ULL), E--38206 La Laguna, Tenerife, Spain}
\altaffiltext{3}{Consejo Superior de Investigaciones Cient\'{\i}ficas, Spain. {\tt amt@iac.es}}
\altaffiltext{4}{National Optical Astronomy Observatory, 950 N. Cherry Avenue, Tucson, AZ  85719}
\altaffiltext{5}{Departamento de F\'{\i}sica Te\'orica, Universidad Aut\'onoma de Madrid, Cantoblanco 28049  Madrid, Spain; \textit{eva.villaver@uam.es}}
\altaffiltext{6}{Instituto de Astronom\'{\i}a-UNAM, Apartado postal 877, Ensenada, 22800 Baja California, M\'exico \textit{ggs@astrosen.unam.mx}}

\begin{abstract}
We present high spatial resolution ($\approx$ 60--90 milliarcseconds) images of the molecular hydrogen emission in the Planetary Nebula (PN) NGC~2346.
The data were acquired during the System Verification of the Gemini Multi-Conjugate Adaptive Optics System + Gemini South Adaptive Optics Imager.
At the distance of NGC~2346, 700~pc, the physical resolution corresponds to $\approx$ 56 AU, which is slightly higher than that an 
[\ion{N}{2}] image of NGC~2346 obtained with \textit{HST}/WFPC2.
With this unprecedented resolution we were able to study in detail the structure of the H$_2$ gas within the nebula for the first time.
We found it to be composed of knots and filaments, which at lower resolution had appeared to be a uniform torus of material. We explain
how the formation of the clumps and filaments in this PN is consistent with a mechanism in which a central hot bubble of nebular gas surrounding
the central star has been depressurized, and the thermal pressure of the photoionized region drives the fragmentation of the swept-up shell.

\end{abstract}


\section{Introduction}

Molecular hydrogen (H$_2$) was first detected in the Galactic planetary nebula (PN) \n2346 by \cite{ZG88}. Thanks in part to its proximity, since then several authors have studied its H$_2$ gas emission in detail (e.g., \citealt{KGMPW94,KAS96,LAT95,VNM99,ARSC01}). They described an apparent structure of clumps and a torus not uncommon in PNe.
Neutral clumps have been detected in PNe in nearby extended objects such as NGC~6543 or NGC~6852 \citep{ML93} and cometary knots seem to be 
detected often in evolved PNe if the resolution allows it (e.g., \citealt{Odell02, HFB02}). However, the formation of these small scale structures is not well understood and different mechanisms have been proposed to explain their formation. Some involve the formation of clumps prior to photoionization  \citep{S98}, while others have suggested that pre-existing high density structures in the interstellar medium (ISM) are responsible for the structures observed later on embedded in the PN shell \citep{APHF12}. The hydrodynamical evolution of those clumps has been studied by \cite{RVCW03} and \cite{APHF12}, and other mechanisms such as their formation in the progenitor asymptotic giant branch (AGB) atmosphere has been ruled out \citep{HM02}. \citet{GSLSMM06} proposed that clumps are the result of the fragmentation of the swept-up shell.

The central star of NGC~2346 is in a close binary system \citep[][hereafter MN81]{MN81} with a measured period
of $\approx16$ days. Given its small orbital separation, it is possible that the system is the remnant of common-envelope
(CE) evolution. However, alternative scenarios for its
formation have also been proposed. \cite{DKW10}, and \cite{KR93} suggested that the system did not evolve through CE, 
rather, it went through mass transfer from an evolved primary with a radiative envelope into the companion, via a thermally unstable Roche Lobe overflow.

The role of binaries in the evolution from the AGB to the PN phase is not well understood. This is mainly because only 40 binary central-stars
of Planetary Nebulae have been studied \citep{DMPFMJ13}. From these only a small fraction are believed to have undergone CE evolution
(e.g., \citealt{CRJ14}, \citealt{MBC13}). In the CE phase the ejection of the envelope is confined closely to the equator (e.g., \citealt{STC98}, \citealt{RT12}): \cite{RT12} found that 90\% of the outflowing material is confined to an angle of $30^{\circ}$ on either side of the equatorial plane.

It this paper we study in detail the structure and gas distribution of H$_2$ in \n2346 using exquisite, high resolution images from observations with the Gemini Multi-Conjugate Adaptive Optics System (GeMS) + Gemini South Adaptive Optics Imager (GSAOI). We also study the properties of the binary central star system.
In section \S2 we describe the observations; Section \S3 includes a description of the molecular hydrogen emission; in \S4 we discuss the stellar progenitor mass; section \S5 gives a general discussion of the results; finally, we summarize our findings and conclusions in \S6.

\section{Observations}

We used the GeMS + GSAOI during System Verification (March 2013). GeMS \citep{GeMS12} comprises multiple deformable mirrors, and uses three Natural Guide
Stars (NGS) and five sodium Laser Guide Stars (LGS). GSAOI is a near infrared (NIR) camera used with GeMS on Gemini South \citep{Pessev13}. GSAOI
provides diffraction limited images in the 0.9$-$2.4 $\mu$m range, using a $2 \times 2$ mosaic Rockwell HAWAII-2RG $2048 \times 2048$ arrays. The
GSAOI field of view (FOV) is 85\arcsec \ $\times$ 85\arcsec\ with a scale of 0.02\arcsec\ pixel$^{-1}$ and a gap between the
arrays of $\approx$ ~2mm (which corresponds to 2.2\arcsec on the sky).

We obtained narrow-band images in the  \H2 2.122 $\mu$m, \brgamma \ 2.166 $\mu$m, and \h212 2.248 $\mu$m filters with good natural seeing 
conditions (0.4--0.5\arcsec).
We used 5 dither positions 
with individual exposure times of 120~s for the \H2 images and 360~s for the Br~$\gamma$ and \h212 images.
As the object is larger than the FOV, adjacent sky frames were taken with the same exposure time.
Exposure time, on-target, was 600~s for the \H2 image, and 1800~s for the \brgamma \ and \h212 images. Data reduction, including distortion correction,
was carried out using the Gemini IRAF\footnote{Image Reduction and Analysis Facility (IRAF) software is distributed by the National Optical
Astronomy Observatories, which is operated by the Association of Universities for Research in Astronomy, Inc., under cooperative agreement with
the National Science Foundation.}
package v1.12, {\it gsaoi}. After the combination of all the images for each band we found the FWHM varies over the whole FOV, between 60 and 90~mas,
with an average value of 80~mas.
We used 2MASS K$_s$ band magnitudes of several field stars to flux calibrate the \H2 images. 

Transmission of the  \H2 2.122 $\mu$m filter is  about 90\% and the width is 0.032 $\mu$m.
A color correction between the center of the K$_s$ band (2.159 $\mu$m), and the center of the \H2 line (2.122 $\mu$m), line was applied, which 
led to an uncertainty of 1\% in the flux calibration. The statistical error in the flux calibration, as measured from different field stars 
is about 2\%.  Thus, altogether, the error in the flux calibration is approximately 2.2\%. 
We obtained an inverse sensitivity for point-sources of $4.94 \pm 0.10 \times 10^{-18}$  \ergcms\ per electron. The pixel area
is $9.5 \times 10^{-15}$~sr, so the inverse sensitivity for extended sources is $5.25 \pm 0.10 \times 10^{-4}$  \ergcms\,${\rm sr}^{-1}$ per electron.


Fig.~\ref{H2} shows the resulting image of the \H2 2.122 $\mu$m emission.
The signal-to-noise (S/N) ratio of the  extended emission in the \brgamma \ and the \h212 images is less than 2$\sigma$ and they are not shown along the paper.
The [\ion{N}{2}] image was obtained with the WFPC2 on board the \textit{HST} (Proposal SM2/ERO--7129). The images were retrieved from the 
Mikulski Archive for Space Telescopes (MAST)\footnote{https://archive.stsci.edu/hst/}.
Individual images were combined in order to remove the cosmic rays using the 
\textit{DrizzlePac} software \citep{Gonzaga12},
with a resulting exposure time of 1280  s.

\section{H$_2$ emission}

The \H2 emission was first detected in \n2346 by \cite{ZG88} using a circular variable filter. \cite{KGMPW94}, \cite{VNM99} and \cite{ARSC01} imaged the \H2 emission in the whole nebula, with a seeing of about 2\farcs5. 
They resolved the extended \H2 emission in both the lobes and what they thought was a central torus. 
However our much higher resolution images show that this feature is not a smooth torus, but an aggregation of a large number of clumps.
\cite{VNM99} also obtained spectra in the K-band, obtaining a ratio of
the \H2/\h212 lines of about 14 in the central part and 4.3 in the lobes, while \cite{ARSC01} found a value of 6.6 for the torus.
Our much better resolution images (see Fig.~\ref{H2}) show that the \H2 emission is in the form of clumps and cometary knots (see Fig.~\ref{clumpnw} and Fig.~\ref{knot}), and are located in both the lobes and the central region. 
Clump sizes vary from 0\farcs16 to 0\farcs34 along the semi-major axis, and projected distances from the central star are from 1\farcs6 to 44\farcs2. 
For the brightest clump we derive  a \H2/\h212 line ratio of 4.7, which is consistent with the value derived by \cite{VNM99} for the lobes and by \cite{ARSC01} for the central region.
From our \H2 image we obtain line surface brightness of $8.8 \times 10^{-4}$ and $1.8 \times 10^{-4}$ \ergcms\, ${\rm sr}^{-1}$ for the brightest and weakest clumps, respectively. 
The highest value is 9 times higher than \cite{VNM99} found.
This is due to the dilution effect of their poor spatial resolution. 
If we degrade our \H2 image to \cite{VNM99} spatial resolution (3\arcsec) we obtain a very similar value of 0.9 $\times$ 10$^{-4}$ \ergcms\ ${\rm sr}^{-1}$ compared with their value of $0.7 \pm2  \times 10^{-4}$ \ergcms ${\rm sr}^{-1}$.
The values that we find for the brightest clumps are 2.5 times higher than the predictions of Model 1 from \cite{VNM99} in the 3000--10000~yr interval, and 5 times higher than the value that \cite{SMJK93} found in 
NGC~6720.
Model 1 in \cite{VNM99} predicts \H2 emission originating in the photodissociation region (PDR), created at the edge of the neutral shell by the UV radiation.  Thus, our values seems to favor shock
excitation. However \cite{VNM99} model reproduced the average surface brightness, and did not explore the full parameter space.


\section{Stellar Progenitor Mass}

The central star of NGC~2346 is a spectroscopic binary (MN81) with a period of 15.99 days. The main sequence companion of the ionizing star, which is seen in Fig.~\ref{H2}, is a 1.8 M$_{\odot}$ A5V star with T$_{eff}$ = 8000~K and L = 18 L$_{\odot}$ (MN81). From the magnitude and extinction of the companion, MN81 obtained a distance of 700 pc to the system. The ultraviolet excess of the ionizing star is compatible with a white dwarf with L = 50 L$_{\odot}$ and T$_{eff}$ = 100,000~K.

The mass of the ionizing star carries the uncertainty in the unknown inclination angle of the binary orbit,
but a reasonable assumption is to consider the orbital plane to be perpendicular to the bipolar lobes of the nebula and zero orbital eccentricity \citep{JLS10}.
Given the 120\degr\ inclination angle of the lobes with respect to the line of sight estimated by \cite{ARSC01}, the binary orbital plane would be seen at an angle of i = 60\degr\ and within $35^\circ < i < 85^\circ$ using a 25\degr\ uncertainty  (\citealt{ARSC01}). 
In Table~\ref{tableruns} we list the solutions for the mass of the ionizing source and the separation of the two components using this plausible range of inclination angles for the binary system.

Using the plausible range of inclination angles, we obtain a mass range for the ionizing source between 0.32 and 0.72 M$_{\odot}$.
The binary separation, which has a very small dependency on the inclination assumed, has values ranging
from 0.159 to 0.167~AU (34.2 $-$ 35.7 R$_{\odot}$). The ionizing star must have evolved past the giant phase,
reaching a radius (either on the red giant or on the AGB) much larger than the current orbital separation of the system.
Thus, it is very likely that the binary system experienced orbital decay if it suffered CE evolution, but significant
orbital decay is not expected if instead mass was simply transferred from the primary when the giant's envelope was not fully convective.
The evolution of a binary system through the CE phase is not understood in detail, and thus it is difficult
to infer the precise history of individual systems (see e.g., \citealt{Iva14}). Given the orbital parameters of the
known components and the mass of the main sequence companion to the PN progenitor, there are two reasonable assumptions
we can make regarding the past history of this system: i) the evolution of the binary did not result in a merger;
and ii) the main sequence mass
of the PN progenitor must have been greater than 1.8 M$_{\odot}$ for it to have evolved into a giant before its companion.

It is difficult to estimate the amount of mass accumulated by the secondary.
Note that the fraction of the envelope that was lost in the CE evolution is unknown. Therefore we can only set a lower limit
to the initial mass of the PN progenitor. Likewise,  we have to assume that the initial orbit separation was larger than 
0.16 AU. But since the spectrum of the ionizing source is unknown, we are unable to determine with certainty whether the CE
phase occurred during the progenitor's red giant branch (RGB) or AGB evolution. However, we argue that the CE phase must have occurred on the AGB, since the orbital separation is expected to decrease during CE evolution \citep{Webb84} and
considering that the largest core mass that can build  up  during the RGB is
$\approx$ 0.47 M$_{\odot}$  (see e.g. \cite{Swei}). This would give the progenitor enough time to build up a
large enough CO core to sustain the photoionization of the nebula. 

\cite{DKW10}, using population synthesis techniques, modeled a CE system to obtain
ZAMS progenitor masses of 2.47 M$_{\odot}$ and 0.98 M$_{\odot}$ and a final WD configuration 
with mass 0.33 M$_{\odot}$.
Such a small WD mass however cannot sustain the ionization of the nebula, and it is not consistent with the UV emission observed. A more massive primary that
experienced enhanced mass-loss before it filled its Roche lobe was suggested by \citet{TE88} for this system.

%

Since the companion has mass of 1.8 M$_{\odot}$, we can set this value as the lower limit to the PN progenitor mass. From \citet{VW93} this value is consistent with L = 50~$_{\odot}$ and T$_{eff}$ = 100,000~K.
Thus, the most plausible solution for the progenitor is a 1.8 M$_{\odot}$ star that has experienced CE evolution on the AGB,
leaving a remnant greater than 0.61 M$_{\odot}$.

We computed the number of ionizing photons for a 1.8 M$_{\odot}$ star assuming a black-body spectrum,
log $L/L_{\odot} = 1.730$, and T$_{eff}$ = 93,540~K, to be $4.0 \times 10^{45}$ photons s$^{-1}$. We will use this result in the modeling described in the next section.

\section{Discussion}
\subsection{Analysis of the Clumps}

\begin{figure}
   \plottwo{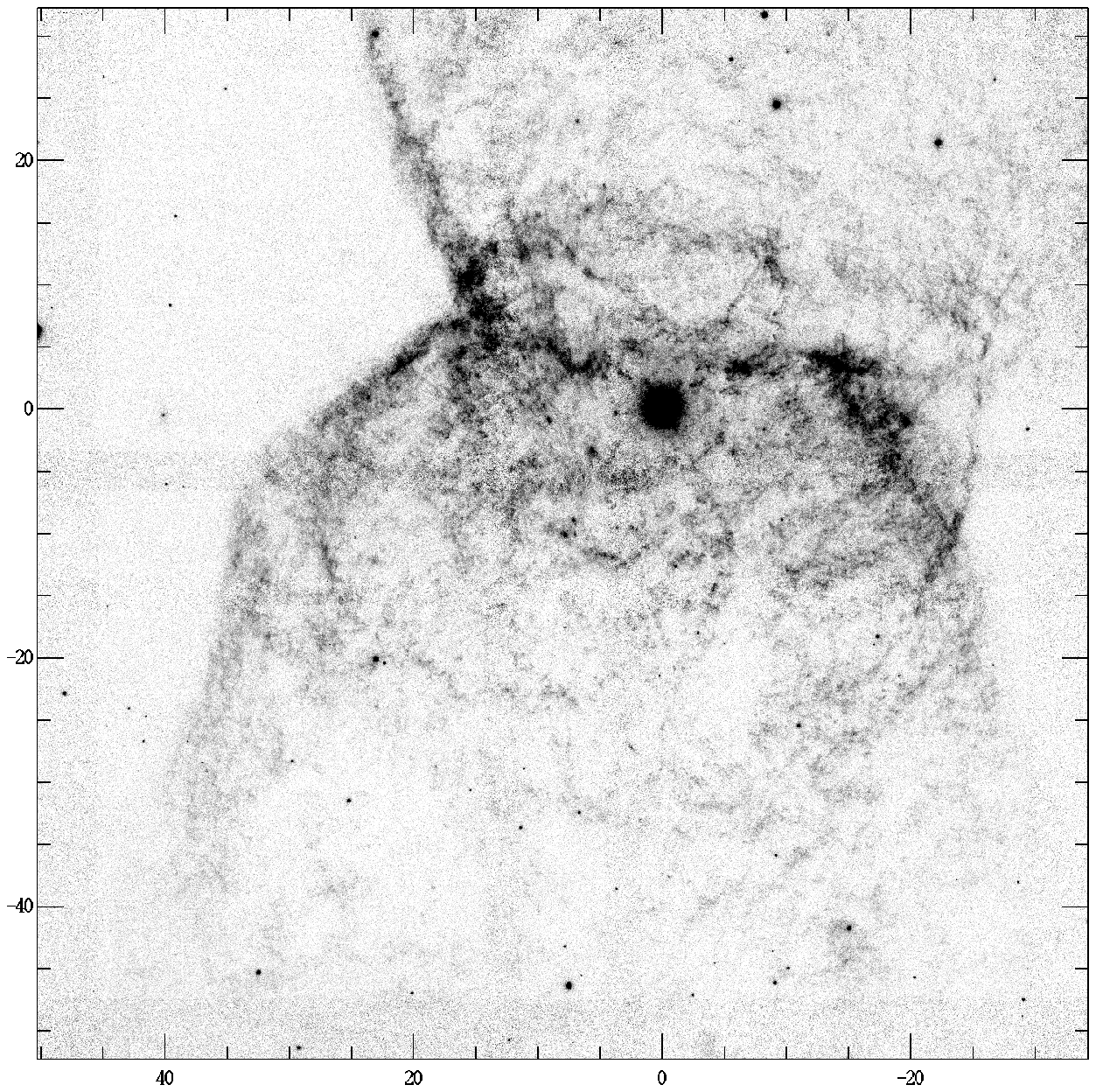}{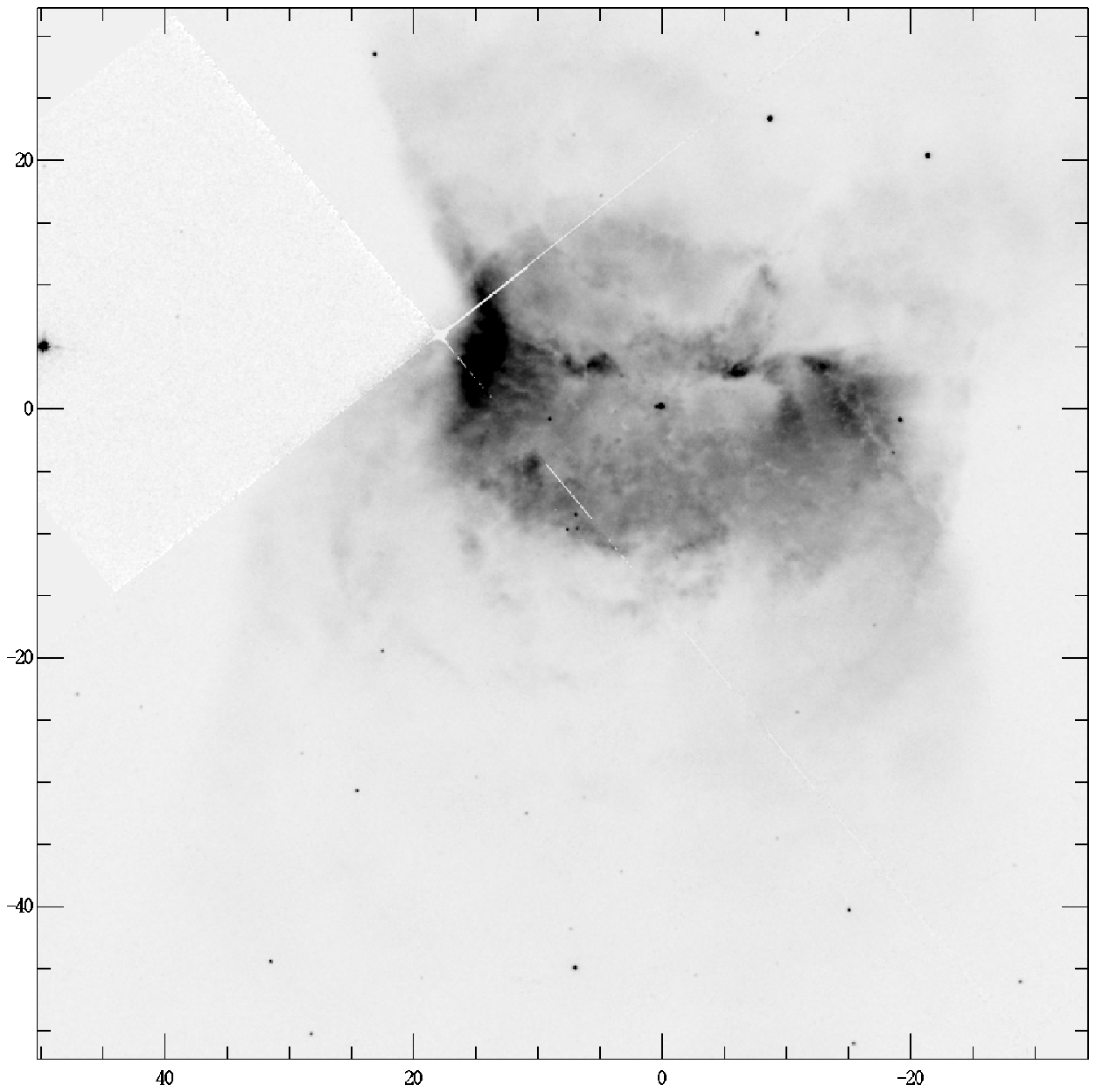}
   \caption[ ]{Images of NGC~2346 obtained with GSAOI in \H2 (\textit{left panel}), and with \textit{HST}/WFPC2 in \nii ~(\textit{right}).
Images are oriented with N up and E to the left; tick labels are in arcseconds with respect to the central star.
   \label{H2}}
\end{figure}

\begin{figure}
   \epsscale{0.5}
   \plotone{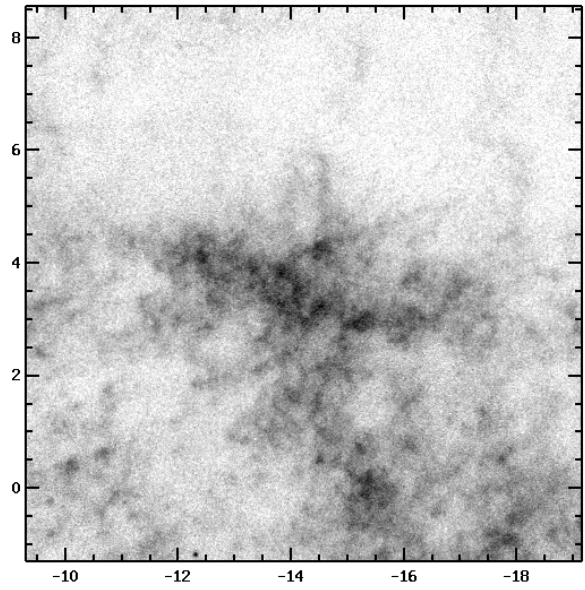}
   \caption[ ]{\H2 clumps NW of the central star.
   \label{clumpnw}}
\end{figure}

\begin{figure}
   \epsscale{0.6}
   \plotone{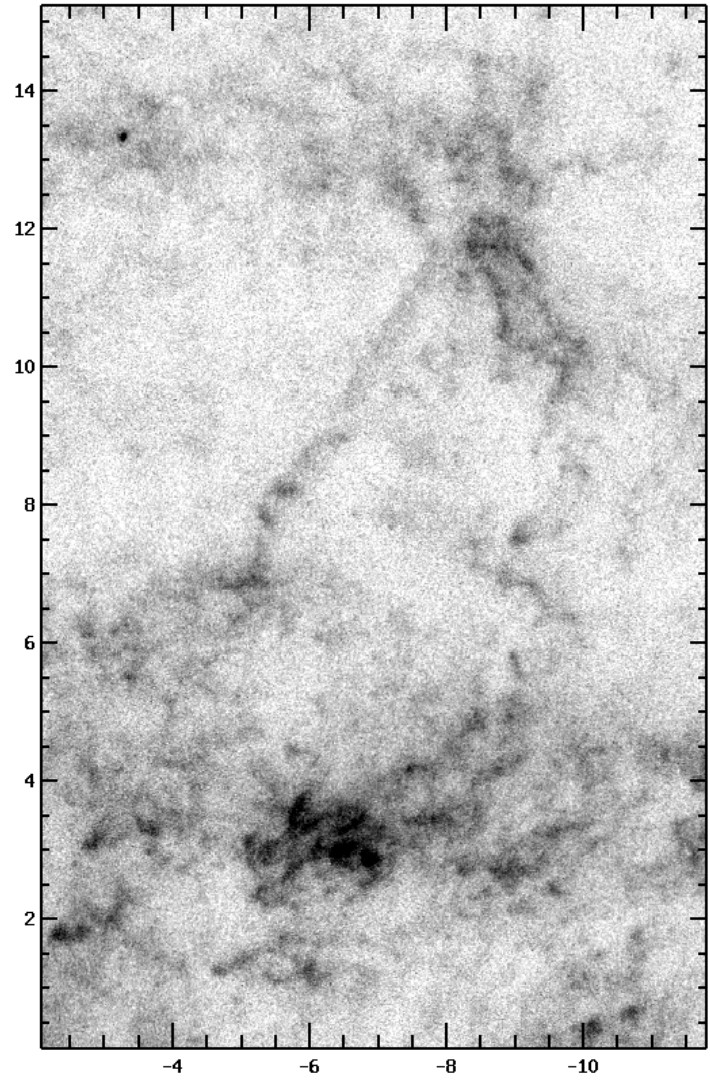}
   \caption[ ]{\H2 cometary knot NW of the central star.
   \label{knot}}
\end{figure}

\begin{figure}
   \epsscale{0.5}
   \plotone{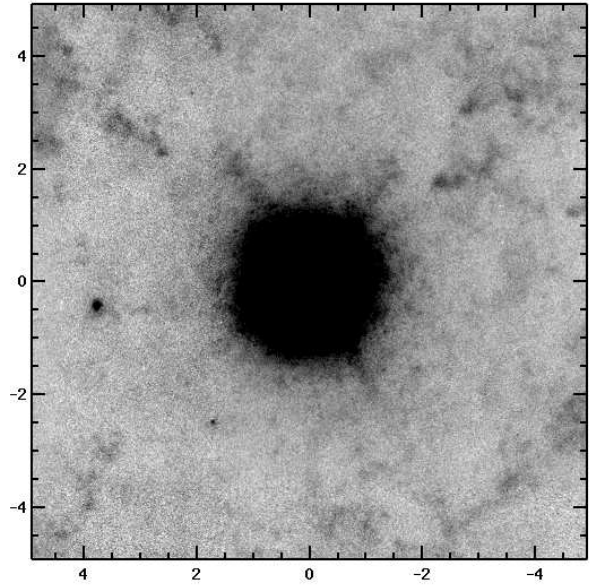}
   \caption[ ]{\H2 cometary knot close to the central star.
   \label{st}}
\end{figure}

\begin{figure}
   \epsscale{0.5}
   \plotone{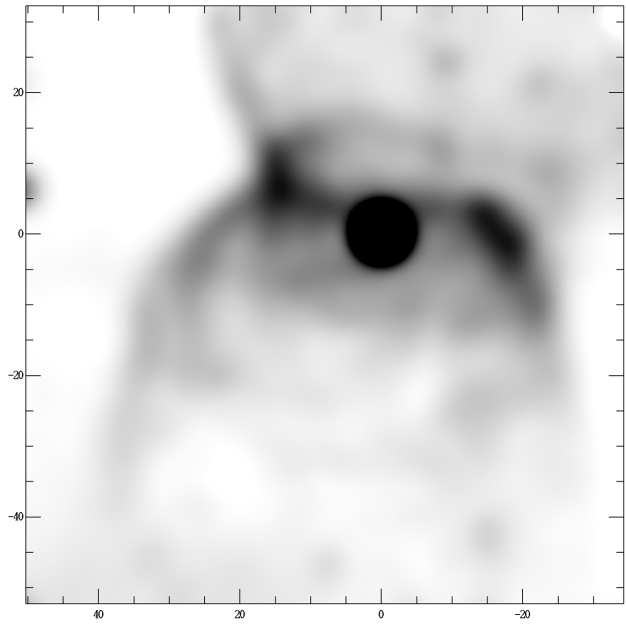}
   \caption[ ]{Same as Figure 1, but convolved with a PSF with a spatial resolution of 2 arcseconds.
   \label{h2g100}}
\end{figure}

As can be seen in Figures~\ref{H2} through \ref{st}, the H$_2$ emission is clumpy.
This clumpy structure is in contrast, and as expected, with the smoother structure seen in the ionized gas (Fig.~\ref{H2}).
Our high resolution images clearly show that the H$_2$ emission is not uniformly distributed in a torus around the central star,
as it has been interpreted from previous, lower resolution observations, but displays a fragmented structure mainly composed of
clumps and cometary knots. To further probe this point we convolved our \H2 image with a symmetric PSF of 2\arcsec,
as shown in Fig.~\ref{h2g100}. It is easy to see how the presence of a torus around the central star might be inferred from a lower resolution image.
Most of the clumps are generally superimposed on a diffuse emission (see Figures~\ref{H2} through \ref{st}).
They are sometimes grouped as in Fig.~\ref{clumpnw}, have cometary tails as in Fig.~\ref{knot}, or have bow shock shape as in Fig.~\ref{st}.
All of this makes it very difficult to measure the number and distribution of clumps in the nebula (e.g., total number of clumps, radial distribution, 
fluxes and knot structure). In any case, the \H2 distribution in \n2346 is completely different than the one in NGC 7293 \citep{MSM09} 

Assuming a distance of 700~pc, each pixel in our image corresponds to a physical size of 14~AU.
From Fig.~\ref{H2} we find that most of the knots are concentrated on the  waist of the nebula (East-West),
and at the edges of the lobes. In the lobes, at distances further than 0.02~pc from the central star, the H$_2$ emission
is more diffuse.
The clumps are at distances of 0.0089 to 0.15~pc from the central star. Typical sizes are in the 1.67--$3.52 \times 10^{15}$~cm (112 to 238~AU) range.

\subsection{Formation of the Clumps}

Several mechanisms have been explored to explain the formation of molecular clumps and cometary tails in PNe. 
A scenario in which pre-existing high density structures in the ISM are responsible for the structures observed \citep{APHF12} is difficult to rule out a priori. 
The question is then how the ISM high density clumps were formed in the first place. 
The mechanism proposed by \cite{GSLSMM06} provides a solution for both the formation of the observed H$_2$ clumps in the PN and the possible existence of long-life high density clumps in the ISM. 
In addition, when high resolution observations of H$_2$  in PNe are obtained, they often reveal clumpy structures \citep{ML13}. 
To assume that the ejected AGB shell would always encounter a clumpy ISM does not seem reasonable; it would imply the ISM to be densely populated by high density structures along many lines of sight. 
We prefer the more plausible explanation offered by \cite{GSLRF99}
that Rayleigh-Taylor instabilities, or the effects of the ionizing radiation in the nebula, lead to the formation 
of clumps. \cite{GSLSMM06} proposed that the cessation of a fast stellar wind can self-consistently produce the type of fragmented 
structures we observed in NGC~2346. The mechanism only requires a rapid decline (i.e., switching off) of the fast stellar wind in 
an ionization-bounded PN. If the stellar wind becomes negligible the hot, shocked bubble depressurizes, and the thermal pressure of the 
photoionized region, at the inner edge of the swept-up shell, becomes dominant. The shell tends to fragment, creating clumps of neutral gas 
with comet-like tails and long, photoionized trails in between, while the photoionized material expands back toward the central star. In the 
hydrodynamical simulations by \cite{GSLSMM06} the cometary globules originate in the neutral, swept-up shell of piled up AGB wind and so they 
must be excited by shocks.

There are two mechanisms arising in very different physical conditions that can be responsible for
the observed H$_2$ emission: (i) shock excitation (\citealt{SH78}),  or (ii) UV pumping in photo-dissociation regions (e.g., \citealt{BD87}).
\cite{ARSC01} have explored the nature of the excitation in NGC~2346 using the relative intensities
of the \H2 and the  \h212 lines, and concluded that the H$_2$ emission line ratios are 
consistent with shock excitation, both in the central nebular regions and in the lobes. 
Further support for shock excitation comes from the incompatibility of \H2 flux with UV excitation, described in $\S$3.

Since the spectrum of the central star is unknown, we cannot determine for certain if the antecedent fast wind has decayed,
leading to depressurization of the hot bubble.  However, there is some evidence of this being the case.
First, the  central star is relatively massive, so the wind would be expected to evolve quickly \citep{VMG02}.
Second, in spite of moderate interstellar extinction, if a robust, fast wind were present one would expect X-ray emission.
Yet NGC~2346 has not been detected with either \textit{XMM} imaging \citep{GRU06} nor \textit{Chandra} observations \citep{KAS12}.
Furthermore, orbital decay is compatible with the nebula kinematical age ($>$ 3,500 yr) derived by \cite{ARSC01}, which, according to \cite{VMG02}, is evidence that the fast wind has already declined below its maximum kinetic energy.

To further examine the formation of the clumps, we have used the 3D version of the \cite{GSLSMM06} model, with AGB wind velocity v$_{exp}$ = 10 \kms,  AGB mass loss $\Mdot$ = 10$^{-5}$ \mlr, 
fast wind velocity v$_{exp}$ = 1000 \kms, and fast wind mass loss $\Mdot$ = 10$^{-7}$ \mlr. 
The temporal grid has $125^3$ zones, and physical dimensions $0.8 \times  0.8 \times  0.8$ pc. 
The fast wind was terminated in the simulation at 1000~yr from the onset.
The photoionization follows the  prescription of \cite{GSF96}, assuming a central ionizing emission of $10^{45}$ s$^{-1}$ ionizing photons (see above). 
To obtain bipolarity, we assumed that the rotation velocity approached  95\% of its critical value, following \cite{GSLRF99}. 
The simulation was done using the hydrodynamical code ZEUS-3D \citep{ZEUS}.
%
It must be noted here that this simulation is highly qualitative, due to the following factors: one, the grid resolution was low, and two, the simulation was not tailored to explain \n2346 in detail but instead
used standard average parameters (although the model was optimized for NGC 2346 in that it assumed a similar number of phoionizing photons as those derived from the NGC 2346's UV flux). Nevertheless, this model is
useful for understanding clump formation.

\begin{figure}
   \epsscale{1.0}
   \plotone{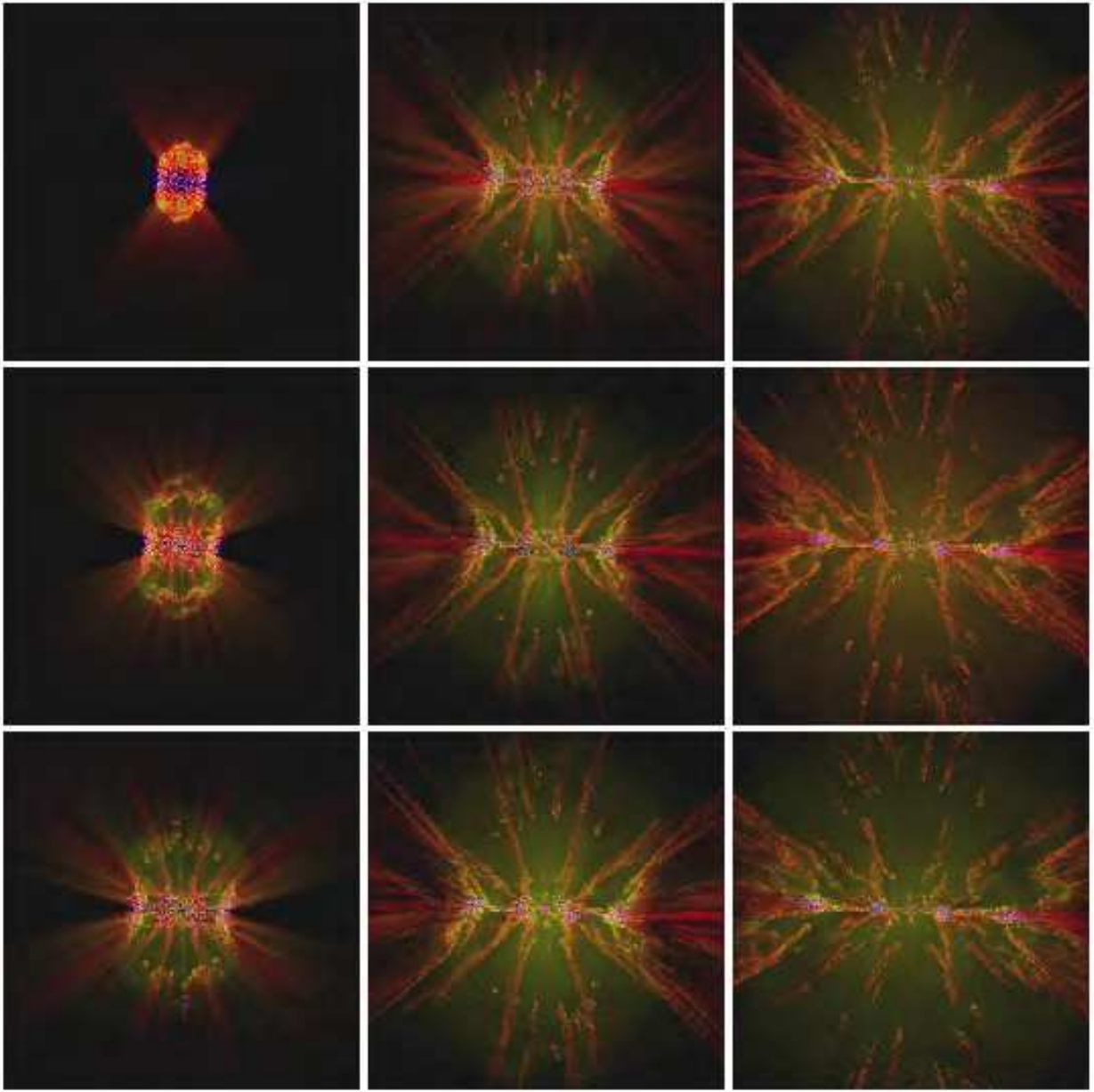}
   \caption[ ]{Snapshots of the emission measure (cm$^{-6}$ pc) along the line sight of the modelled gas.
Three different temperatures are represented in  different colors (see text for details).
Starting at 1000~yr elapsed time with the cessation of the stellar wind (\textit{upper left}), the evolution proceeds from upper to lower, then left-to-right, with a timestep of 1000~yr.
   \label{tex9}}
\end{figure}

In Fig.~\ref{tex9} we show the gas emission measure for nine model snapshots, on the X-Z plane. The initial model is for t=1000~yr (following the termination of the fast stellar wind), and 
subsequent 
models are at timestep of 1,000 yr, covering 9,000 yr of the evolution. The emission of the various components are represented by different colors: photoionized gas with temperature$\sim$10,000 K is represented in green;
 gas (either photo ionized or neutral) with temperature above 1,000 K is in red,  and gas with temperature below 1,000 K (blue). The blue (cool) gas emission follows the molecular emission (i.e., H$_2$), 
so we can use this latter indicator to see what happens in the molecular regime.
The simulation shows, as a general view, an initial toroid of cool gas at the equator. Once the swept-up shell is highly fragmented, the toroid is no longer visible and only the large clumps, with an optical  depth big enough to shield the ionizing radiation, will survive and be detected.

The current evolutionary stage of \n2346 is represented by models of times between the first and the second snapshot of Fig.~\ref{tex9}, since it appears that the
clumps in \n2346 are just being formed, and the photoionization front has not yet reached the equatorial latitudes. \n2346 has a small 
number of long cometary knots (Fig.~\ref{knot}), not easily recognable in the \textit{HST} \nii ~image (Fig.~\ref{H2}).
This contrasts with the large number of such knots in the Helix (NGC~7293) or in the Eskimo (NGC~2392) nebulae \citep{Odell02}.
In addition the size of the clumps in NGC~7293 (94~AU) are similar in size to those in \n2346 \citep{Odell07}.

%

\subsection{The Survival of H$_2$ Clumps}

An interesting question is to determine how the molecular hydrogen clumps can survive the photoionization front. If the propagation velocity $dR/dt$ of the ionization front within a clump can be expressed according
to equation 7.5 of \cite{DW97}, then

\beq \frac{dR}{dt} = \frac{S_*}{4\pi\,R^2\,n_0}
-\frac{1}{3}\,Rn_0\,\beta_2,  \eeq

\noindent where $S_{*}$ is the number of photoionizing photons, $R$ the distance from the ionizing source to the clump, $dR$ the size of the clump,
$\beta_{2}$ the recombination coefficient, and $n_0$ the density of the clumps.
For \n2346, the closest clumps (see Fig.~\ref{st}) are at a projected distance of $2.75 \times 10^{16}$~cm (1850 AU)
from the central star, with size $\sim2.1 \times 10^{15}$~cm. Thus, assuming
$S_{*} = 4.0 \times 10^{45}$ s$^{-1}$
and the kinematical age as a proxy for the evolutionary time (3500~yr), we estimate the minimum density
required for clumps to survive the passage of the ionization front. We obtain a value of $n_0 = 15,000$ cm$^{-3}$,
which is smaller than that of the gas density derived for the largest clumps ($5 \times 10^4$ cm$^{-3}$) from dust
absorption of the [\ion{O}{3}] images in NGC~6853 by \cite{ML93}.

During post-AGB evolution the stellar luminosity declines, and some nebular clumps may not be photoionized. 
Instead, they survive as molecular material, eventually making the ISM clumpy. 
Since the \H2 will not couple with the 3~K cosmic microwave background, the only destruction mechanism in the ISM 
is the spontaneous radiative dissociation \citep{SD71}. 
\cite{PCM94} proposed that H$_2$ clumps of a size of about 30 AU can explain baryonic dark matter in spiral galaxies. 
The H$_2$ clumps in \n2346, with sizes are between 112--238 AU, will populate the ISM, contributing to the baryonic 
dark matter of the Galaxy. Whether \H2 in clumps survive or not past the PN stage and populate the ISM of galaxies 
is beyond the scope of this work. We speculate that if clumps survive, given the amount of mass they contain, they should 
be accounted for as a source of baryonic dark matter.

\section{Conclusions}

We have obtained molecular hydrogen (H$_2$) images of \n2346 with unprecedented spatial resolution: 56~AU at an adopted distance of 700~pc.
The images reveal the H$_2$ emission to be fragmented in clumps and cometary knots, rather than a uniform disk as previously thought.
The clumps range in size from 112 to 238 AU; the clump apparently closest to the central star lies at a projected distance of 1850~AU.
The central star has undergone binary interaction with its nearby companion, probably during its AGB phase.
The ionizing star has a minimum mass of 0.61 M$_{\odot}$, and an ionizing flux of $4.0 \times 10^{45}$ photons s$^{-1}$.

We performed a hydrodynamical simulation that shows how an initial disk or toroid breaks up into individual clumps, once the swept-up shell 
fragments following the decline of the fast wind. In order to survive the ionization front, the pre-ionization density of the molecular 
hydrogen clumps must exceed $\sim15,000$ cm$^{-3}$, which is found in some PNe. Those clumps that survive the ionization front will 
eventually populate the ISM, and may contribute to the baryonic dark matter in our Galaxy.

\acknowledgments

We are grateful to R. Carrasco for the data reduction of the GSAOI images. We thank Wolfgang Steffen and Dolores Bello for fruitful discussion.
A.M. and D.A.G.H. acknowledge support for this work provided by the Spanish Ministry of Economy and Competitiveness under grant 
AYA-2011-27754. L.S. and R.A.S. acknowledge support for this project from NOAO.
E.V. acknowledges support from grant AYA 2013-45347P.
G.G.-S. is partially supported by CONACyT grant 178253 and DGAPA grant IN100410.
G.G.-S. thanks Michael L. Norman and the Laboratory for Computational Astrophysics for the use of ZEUS-3D.
Some of the data presented in this paper were obtained from the Mikulski Archive for Space Telescopes (MAST). STScI is operated by 
the Association of Universities for Research in Astronomy, Inc., under NASA contract NAS5-26555. Support for MAST for non-HST data is provided 
by the NASA Office of Space Science via grant NNX09AF08G and by other grants and contracts.
Based on observations obtained at the Gemini Observatory, which is operated by the 
Association of Universities for Research in Astronomy, Inc., under a cooperative agreement 
with the NSF on behalf of the Gemini partnership: the National Science Foundation 
(United States), the National Research Council (Canada), CONICYT (Chile), the Australian 
Research Council (Australia), Minist\'{e}rio da Ci\^{e}ncia, Tecnologia e Inova\c{c}\~{a}o 
(Brazil) and Ministerio de Ciencia, Tecnolog\'{i}a e Innovaci\'{o}n Productiva (Argentina).

\begin{deluxetable}{ccc}
\tabletypesize{}
\tablecolumns{3}
\tablewidth{0pt}
\tablecaption{Masses and separation of the binary system \label{tableruns}}
\tablewidth{0pt}

\tablehead{
\colhead{Inclination Angle} & \colhead{Progenitor Mass}  & \colhead{Separation} \\
\colhead{(deg)} & \colhead{(M$_{\odot}$)}  & \colhead{(AU)}
}
\startdata
30 & 0.72 & 0.167 \\
35 & 0.61 & 0.166 \\
38 & 0.56 & 0.165 \\
40 & 0.53 & 0.164 \\
50 & 0.43 & 0.162 \\
60 & 0.37 & 0.161 \\
70 & 0.34 & 0.160 \\
80 & 0.33 & 0.159 \\
85 & 0.32 & 0.159 \\

\enddata
\end{deluxetable}

\end{document}